\documentclass[twocolumn]{aastex62}
\submitjournal{ApJ}
\usepackage{enumitem}
\usepackage{graphicx,url}
\usepackage{color}
\usepackage{amsmath}
\def\gx339{GX~339$-$4}
\def\jj1550{XTE~J1550$-$564}

\def\xmm{{\it XMM-Newton}}

\def\nustar{{\it NuSTAR}}

\def\nicer{{\it NICER}}
\def\xstar{{\tt xstar}}
\def\reflionx{{\tt reflionx}}
\def\reflionxBB{{\tt reflionx\_BB}}
\def\BBrefl{{\tt BBrefl}}
\def\xillver{{\tt xillver}}

\def\xillverNS{{\tt xillverNS}}
\def\relxill{{\tt relxill}}

\def\relxillCp{{\tt relxillCp}}

\def\relxillNS{{\tt relxillNS}}

\def\relconv{{\tt relconv}}

\def\rfxconv{{\tt rfxconv}}
\def\xilconv{{\tt xilconv}}
\def\xspec{{\sc xspec}}
\def\isis{{\sc isis}}
\def\sherpa{{\sc sherpa}}
\def\spex{{\sc spex}}
\def\Ga{$\Gamma$}
\def\Ecut{$E_\mathrm{cut}$}
\def\kTbb{$kT_\mathrm{bb}$}

\def\logXi{$\log(\xi/$erg\,cm\,s$^{-1})$}
\def\Afe{$A_\mathrm{Fe}$}
\def\logne{$\log(n_e/$cm$^{-3})$}

\def\deg{$^{\circ}$}
\shorttitle{X-ray Reflection Models for Neutron Stars}
\shortauthors{Garc\'{\i}a \& et al.}

\begin{document}

%\title{\large\bf Accurate Treatment of Comptonization in X-ray Reflection Models}
\title{\large\bf Relativistic X-ray Reflection Models for Accreting Neutron Stars}
%\note{(Maybe think of a better title?)}

\correspondingauthor{Javier~A.~Garc\'ia}
\email{javier@caltech.edu}

\author[0000-0003-3828-2448]{Javier~A.~Garc\'ia}
\affil{Cahill Center for Astronomy and Astrophysics, California Institute of Technology, Pasadena, CA 91125, USA}
\affil{Dr. Karl Remeis-Observatory and Erlangen Centre for Astroparticle Physics, Sternwartstr.~7, 96049 Bamberg, Germany}

\author{Thomas~Dauser}
\affil{Dr. Karl Remeis-Observatory and Erlangen Centre for Astroparticle Physics, Sternwartstr.~7, 96049 Bamberg, Germany}

\author[0000-0002-8961-939X]{Renee~Ludlam}\thanks{NASA Einstein Fellow}
\affil{Cahill Center for Astronomy and Astrophysics, California Institute of Technology, Pasadena, CA 91125, USA}

\author{Michael~Parker}
\affil{Institute of Astronomy, Madingley Road, Cambridge CB3 0HA, UK}

\author{Andrew~Fabian}
\affil{Institute of Astronomy, Madingley Road, Cambridge CB3 0HA, UK}

\author{Fiona~A.~Harrison}
\affil{Cahill Center for Astronomy and Astrophysics, California Institute of Technology, Pasadena, CA 91125, USA}

\author{J\"orn~Wilms}
\affil{Dr. Karl Remeis-Observatory and Erlangen Centre for Astroparticle Physics, Sternwartstr.~7, 96049 Bamberg, Germany}

%==================================================================================
\begin{abstract}
We present new reflection models specifically tailored to model the X-ray
radiation reprocessed in accretion disks  around neutron stars, in which the
primary continuum is characterized by a single temperature blackbody spectrum,
emitted either at the surface of the star, or at the boundary layer. These
models differ significantly from those with a standard power-law continuum,
typically observed in most accreting black holes.  We show comparisons with
earlier reflection models, and test their performance in the \nustar\
observation of the neutron star 4U~1705$-$44. Simulations of upcoming missions
such as {\it XRISM-Resolve} and {\it Athena} X-IFU are shown to highly the
diagnostic potential of these models for high-resolution X-ray reflection
spectroscopy.  These new reflection models \xillverNS, and their relativistic
counterpart \relxillNS, are made publicly available to the community as an
additional flavor in the {\sc relxill} suite of reflection models.

\end{abstract}

\keywords{accretion, accretion disks -- atomic processes -- neutron stars
-- line: formation -- X-rays: individual (4U~1705$-$44)}

%
%==================================================================================
%
%
%
\section{Introduction}\label{sec:intro}

Studying X-ray reflection of high-energy photons emitted near a compact object
and then reprocessed and reflected off an accretion disk, has traditionally
been focused in the case of accreting black holes, both of stellar-mass in
binary systems and supermassive black holes in active galactic nuclei. This is
because the physics of the problem is essentially independent of the mass of
the central object.  For this same reason, this technique, normally referred to
as X-ray reflection spectroscopy, can be equally applied to accreting neutron
star systems \citep[e.g.;][]{bah07,cac08}.

There exist, however, some differences between accreting black holes and
neutron stars.  The most obvious and relevant in the context of X-ray
reflection is the fact that in a black hole system the accretion disk is
expected to be sharply truncated at the inner-most circular orbit (ISCO),
whereas in a neutron star system the inner edge of the disk can reach the
surface of the star if not impeded by a boundary layer \citep{pop01,dai10}, or
the magnetosphere \cite[e.g.;][]{ibr09,cac09,pap09,lud17}.  In particular,
\cite{dai10} showed that the primary continuum during the soft state of
4U~1705$-$44 can be attributed to the boundary layer and modelled the
reflection component accordingly \citep[see also][]{egr13,pin15,chi16}.

Additionally, the structure of the disk can also be different, as the simple
corona-disk model of \cite{sve94} predicts much higher density for disks around
neutron stars than for those around black holes, unless the accretion rate is
very low.  Furthermore, while the relativistic effects such as boosting and
gravitational shifts of the spectral features are independent of the nature of
the compact object, neutron stars have much lower spin values than black holes.
Due to this slower rotation speed \citep[$a_*<0.3$, e.g.;][]{gal08,mil11}, and
the solid surface of the neutron star, the inner edge of the disk is at a much
larger radius than for a rapidly rotating black hole. As the strength of the
relativistic distortion strongly increases with smaller radius \citep[see,
e.g., ][]{dau10}, reflection from the innermost region of the accretion disk
around a neutron star will therefore show a much more subtle distortion due to
relativistic effects in comparison to black holes typically with a high spin.
Such a small spin value also means the reflection will not be a dominant
component in the spectrum, due to the much lower reflection fraction
\citep{dau14}.  Moreover, neutron stars must radiate all the accretion energy,
whereas black holes swallow some (i.e orbital energy). Also important for
reflection spectroscopy is the fact that the metric describing the space-time
around a neutron star can differ significantly from the Kerr metric for black
holes, particularly for neutron stars with large angular momentum
\citep{sib98}. 

The relativistic blurring effects mentioned above are the main tool used in
reflection spectroscopy to access physical information of the central object,
such as estimates of the spin, location of the inner accretion disk, and
inclination of the system. This is done through careful modeling of atomic
features in the reflection spectrum, with the most prominent being that due to
the iron emission complex near $6.4-7.1$\,keV.  Broad iron lines (as well as
other lines at soft energies) have been observed in several neutron star
binaries \citep[e.g.;][]{bah07,dis09,cac10,cac12}, and more recently thanks to
the improved sensitivity and bandpass of the instruments onboard of the
\xmm, \nustar\ and \nicer\ X-ray observatories
\citep[e.g.;][]{mil13,deg15,deg16,mat17, mon18,jai19,mon20,van20,kol21}.

Models to treat X-ray reflection have been in continuous development for the
past several decades.  For a comprehensive review of the literature, the reader is
referred to \cite{fab10,gar13a,dau16b}. Primitive models only considered the
scattering problem in a optically-thick medium in a fully neutral gas
\citep[e.g.;][]{lig81,lig88}, in some cases including emission K lines from
iron \citep[e.g.;][]{mat91}. Later, the importance of other astrophysically
abundant elements was recognized as major contributors to the observed features
\citep[e.g.;][]{rey97}. A consistent treatment of the ionization balance in an
X-ray illuminated slab was first developed by \cite{ros93}, producing the
widely used reflection model \reflionx\ \citep{ros05}. This code was used to
compute reflection models tailored for neutron stars by taking into account the
appropriate parameters for the illumination \citep{ros07,bal04}. However, the
ionization structure in these models is somewhat limited by the use of outdated
atomic data for the relevant transitions.

A major advance in reflection models was introduced in the \xillver\ model by
\cite{gar10}, which includes the largest and most recent atomic database for
inner-shell transitions, implementing the routines from the photoionization
code \xstar\ for the determination of the ionization and energy balance. The
code \xillver\ also provides a more accurate radiative transfer calculation
based on the Feautrier method as described in \citep{mih78}, which includes a
fully angle-dependent solution for the reflected spectrum \citep{gar13a}.
Furthermore, pre-computed \xillver\ spectra are self-consistently linked to the
relativistic blurring convolution code \relconv\ \citep{dau10,dau13}, which is
known as the suite of models \relxill\ \citep{gar14a,dau14}.

In this paper we present a new flavor of the \relxill\ models specifically
tailored to describe the X-ray radiation reprocessed in accretion disks around
neutron stars, in which the primary continuum is characterized and dominated by
a blackbody spectrum, rather than the standard power-law continuum typically
observed in most accreting black holes. Preliminary versions of these models,
referred to as \relxillNS, have been already tested and implemented to analyze
data for several neutron star X-ray binaries observed with the \nustar\ and
\nicer\ observatories \citep[e.g., Serpens~X-1, GX~3+1, and
4U~1735$-$44;][respectively]{lud18,lud19,lud20}.

We also present comparisons with the standard models based on power-law
illumination, and with previously calculated reflection models that also
consider a blackbody illumination. An example of the performance of these
models is shown for the case of the neutron star system 4U~1705$-$44.  Like
other flavors of the \relxill\ suite, these models are made publicly available
to the community for they use in any of the traditional X-ray fitting packages,
such as \xspec\ \citep{arn96}, \isis\ \citep{hou00},
\sherpa\ \citep{fre01}, and \spex\ \citep{kaa96}.
%

%----------------------------------------------------------------------------------
\section{The X-ray Reflection Model}\label{sec:model}

To compute the reprocessed (reflected) X-ray spectrum out of an illuminated
accretion disk around a neutron star, we make use of a modified version of our
reflection code \xillver. In this case, the illuminating radiation field is
changed from the standard power-law spectrum (mostly appropriate for sources
in the hard state), to a spectrum described by a blackbody radiation field at a
given temperature, appropriate for sources with such a continuum, as it is the
case of some accreting neutron stars.  The implicit assumption is that in these
systems most of the X-ray continuum emission is thermal radiation originating
from either the surface of the neutron star (whether that be uniform thermal
emission or a hot spot on the surface), or from the boundary layer region
extending from the surface of the star.  This continuum radiation then
illuminates the disk and produces a reflection spectrum accordingly. We thus
refer to these new models as \xillverNS.

The \xillverNS\ code solves the radiation transfer problem in a plane-parallel
slab implementing the Feautrier method \citep{fea64,mih78} with two boundary
conditions (at the top and bottom of the slab), in an iterative process. This
solution is coupled with the calculation of the ionization and energy balance
equation making use of the routines from the \xstar\ photoionization code
\citep{kal01}, with the most up-to-date atomic database. The code provides the
angle-dependent spectrum emergent at the top of the slab for a given
irradiation, gas density, and elemental abundances.  An extensive and detailed
discussion on these reflection calculations can be found in
\cite{gar10,gar13a,gar14a,gar16b}.
%

%......................................................................................
\begin{figure*}[ht]
\centering
\includegraphics[width=\linewidth]{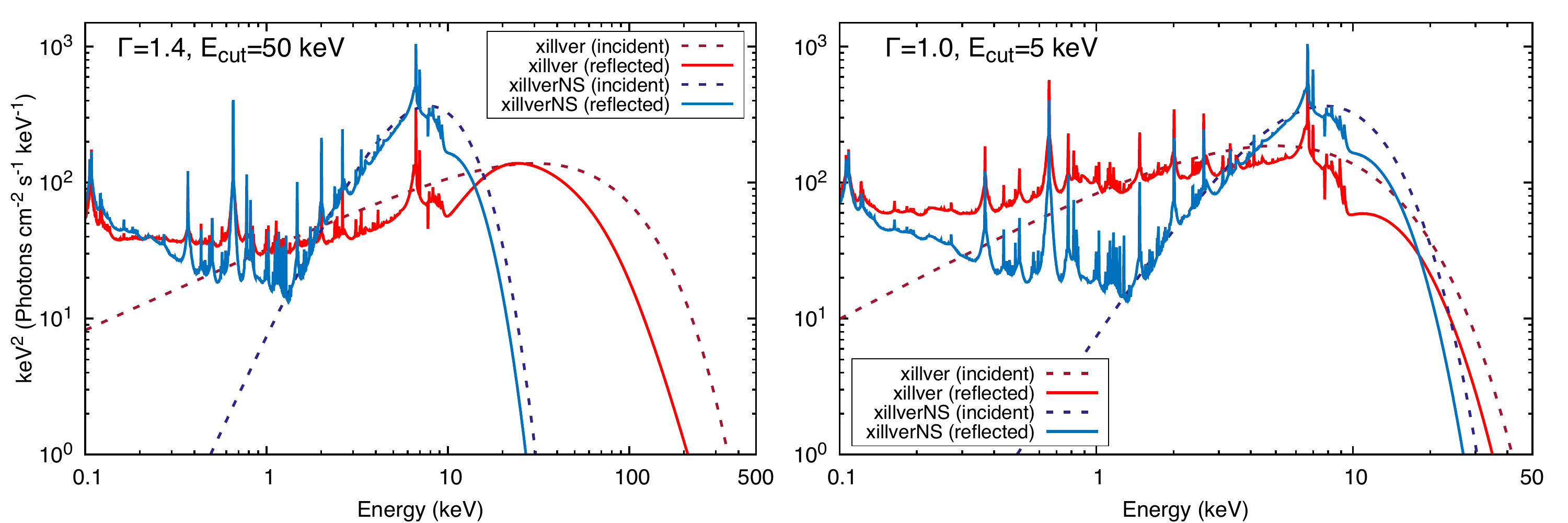}
\caption{
Comparison of the reflection calculations produced with the \xillver\ code
using the standard power-law illumination (red), and those produced with the
new model \xillverNS\ implementing a blackbody illumination spectrum (blue).
The left panel shows the \xillver\ case for \Ga$=1.4$ and \Ecut$=50$\,keV;
while the right panel shows the extreme case of \Ga$=1.0$ and \Ecut$=5$\,keV.
In both cases, the \xillverNS\ is calculated for \kTbb$=2$\,keV. Other
parameters common to all models are: \logXi$=3.1$, \Afe$=1$, $i=30$\deg, and
\logne$=15$. The reflection fraction was set to $R_{\rm frac}=0$ when plotting
the incident components, and to $R_{\rm frac}=-1$ when plotting the reflected
components.
}
\label{fig:comp_xill}
\end{figure*}
%......................................................................................
%

Finally, and in the same fashion of all our previous models, we produce a grid
of synthetic spectra for a given set of input parameters, each covering a range
of values appropriate for the astrophysical sources of interest.  This model
table is then self-consistently connected with our relativistic convolution
model \relconv\ \citep{dau10,dau13}. The model takes into account the angular
distribution of the solution provided by \xillverNS\ and correctly predicts the
integrated reflection off the disk by including all relativistic effects. This
includes the distortion of the spectral features due to relativistic
corrections such as boosting, gravitational redshift and Doppler effects. As
described in \cite{gar14a}, this procedure is significantly different from
simple convolution of the reflection spectrum, since a given line of sight will
receive contributions from photons emitted at various angles due to light
bending effects.  This complete relativistic model is then referred to as
\relxillNS.

In the following, we describe the main parameters of the model, highlight the
main differences with respect to reflection out of power-law illumination, and
show detailed comparisons with previous calculations reported in the
literature.
%

%----------------------------------------------------------------------------------
\section{Results}\label{sec:results}

\subsection{X-ray Reflection in the Disk's Frame}\label{sec:xillverNS}

In this Section  we describe the calculation of a new grid of reflection
spectra in which the radiation field that illuminates the accretion disk takes
the form of a blackbody spectrum. The resulting spectra are produced for a
single slab at constant density, thus they represent reflection in the frame of
the disk (i.e., ignoring relativistic effects). This new set of models are
designed to be applicable for the case of irradiated accretion disks around
neutron stars. The blackbody X-ray continuum is produced at the surface of the
neutron star, or close to the surface from a boundary layer region, with
typical temperatures in the $\sim0.5-10$\,keV range.  

The shape of the illuminating radiation has a direct effect in controlling the
ionization state and temperature profile of the atmosphere, and thus in
determining the overall shape and spectral features of the reprocessed
radiation. This effect is shown in Figure~\ref{fig:comp_xill}, where we compare
the \xillverNS\ calculations using a blackbody spectrum at a temperature of
\kTbb$=2$\,keV, with the standard \xillver\ spectra assuming a power-law
illumination.  The left panel shows a case representative of power-law
reflection in black hole binaries, i.e., \Ga$=1.4$, and \Ecut$=50$\,keV.  This
spectrum shows a close resemblance to the \xillverNS\ at energies below $\sim
2$\,keV, but at higher energies the differences are evident. The power-law
illumination has a larger number of photons at high energies, which contributes
to a more pronounced Compton hump. Even when the parameters of the power-law
are set to their extreme values in order to produce a spectrum closer to the
blackbody in \xillverNS, i.e., \Ga$=1$ and \Ecut$=5$\,keV (right panel,
Figure~\ref{fig:comp_xill}), the differences in the resulting reflected spectra
are large enough such that they will likely affect the spectral fitting to
observational data.
%

%......................................................................................
\begin{figure*}[ht]
\centering
\includegraphics[width=\linewidth,trim={3cm 2cm 3cm 0.5cm}]{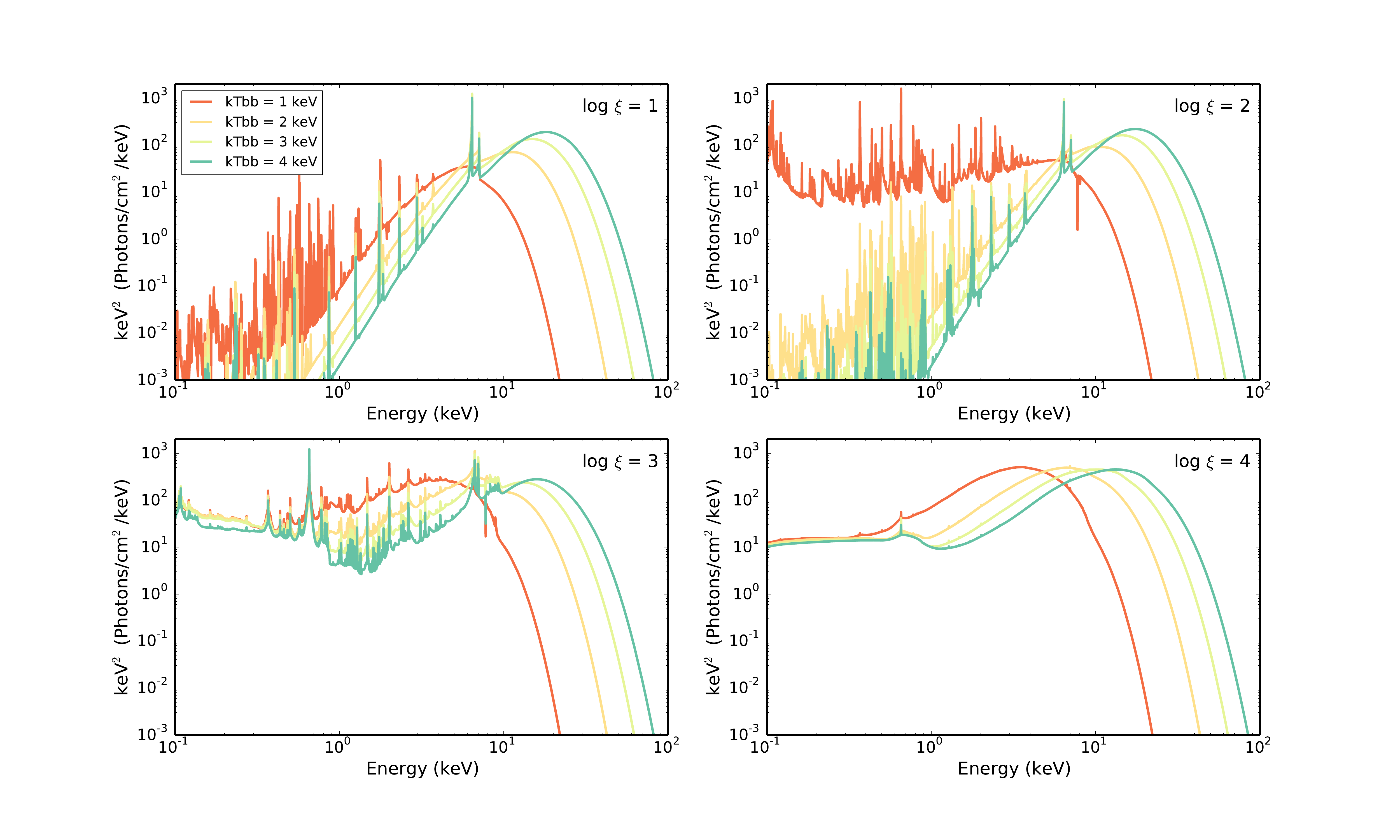}
\caption{
Effects of changing the temperature and overall flux of the illuminating
radiation in the reflection spectra computed with the new \xillverNS\ code.
Each panel shows the reflected spectra for an increasing temperature \kTbb\ of
the blackbody illumination at a given ionization parameter \logXi, as
indicated. Other parameters common to all the models shown are: \Afe$=1$,
$i=30$\deg, \logne$=15$, and $R_{\rm frac}=-1$ (i.e., only the reflection without the
continuum spectrum is shown).
}
\label{fig:xillNS_kTbb}
\end{figure*}
%......................................................................................
%

Given the relevance of the illuminating radiation in shaping the reprocessed
spectrum, we have produced a full grid of reflection spectra using exclusively
a blackbody at a given temperature \kTbb. As in our previous models
\citep{gar10,gar11,gar13a}, \xillverNS\ assumes an illuminating spectrum
incident at 45\deg\ on the surface of a plane-parallel slab with constant
density $n_e$. The slab has a total optical depth of 10, with no illumination
from the bottom. The abundance of all astrophysically relevant elements is set
to their Solar values based on the \cite{gre98} standard, with iron set at
different values \Afe. The net incident flux (integrated in the $0.1-10^3$\,keV
range), is set to match a desired ionization parameter $\xi = 4\pi
F_\mathrm{x}/n_e^2$, for a given blackbody temperature \kTbb\ and density
$n_e$. The complete set of parameters and the values used to produce the final
grid of models is listed in Table~\ref{tab:xillverNS}.
%

%..................................................................................
%
\begin{deluxetable}{ccc}[ht!]
\tabletypesize{\scriptsize}
\tablecaption{List of Parameters for the \xillverNS\ Model \label{tab:xillverNS}}
\tablecolumns{3}
\tablewidth{0pt}
\tablehead{
\colhead{Parameter} & \colhead{Symbol (Units)} & \colhead{Range}
}
\startdata
Blackbody Temperature   & \kTbb\ (keV)  & $[0.5,10]$ \\
Ionization Parameter    & \logXi        & $[1,4]$    \\
Electron Number Density & \logne        & $[15,19]$  \\
Iron Abundance          & \Afe\ (Solar) & $[0.5,10]$ \\
\hline
\enddata
\end{deluxetable}
%
%..................................................................................
%

\begin{figure*}[ht]
\centering
\includegraphics[width=\linewidth,trim={3cm 2cm 3cm 0.5cm}]{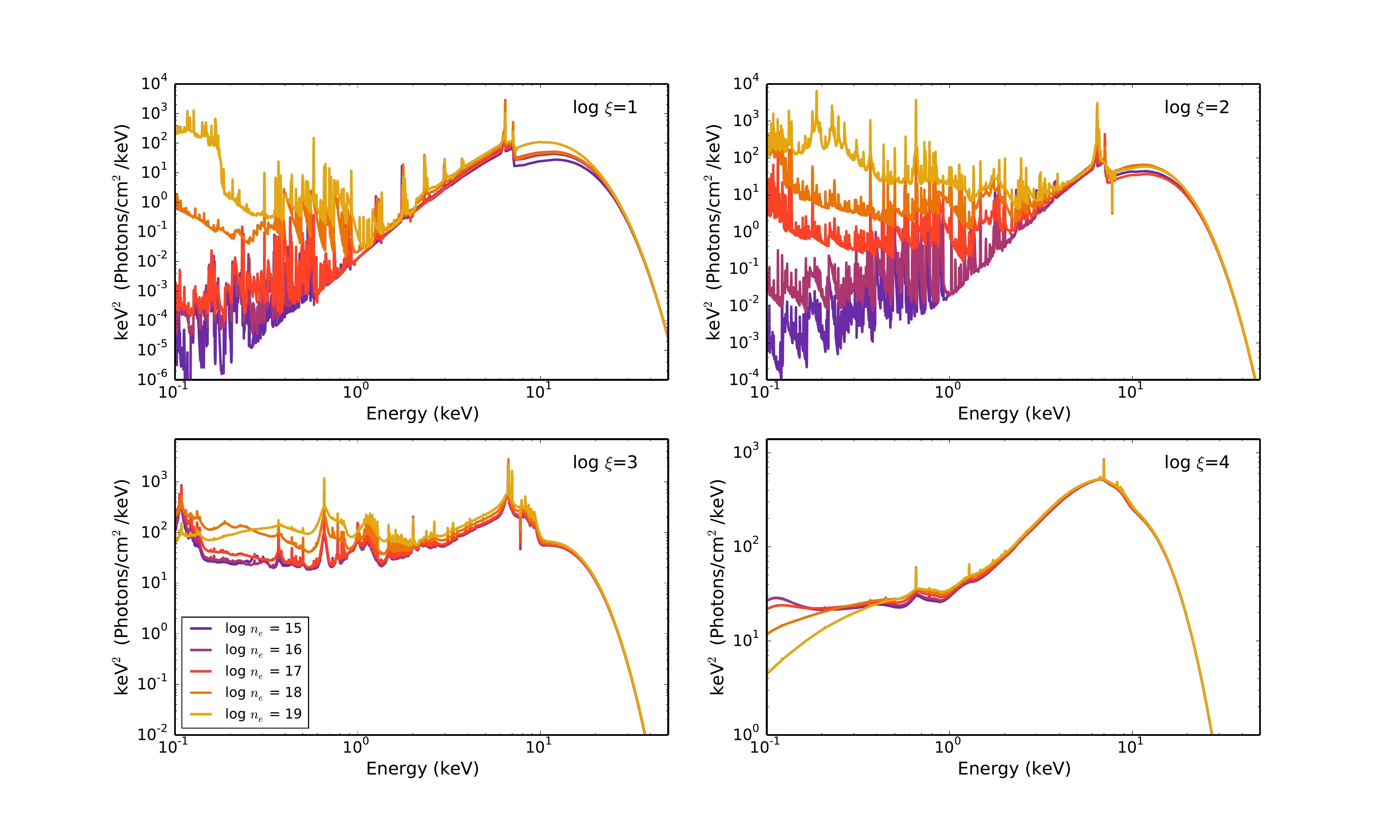}
\caption{
Effects of changing the gas electron density for different illuminating fluxes
in the reflection spectra computed with the new \xillverNS\ code.  Each panel
shows the reflected spectra for increasing density \logne\, at a given
ionization parameter \logXi, as indicated. Other parameters common to all the
models shown are: \Afe$=5$, $i=30$\deg, \kTbb$=2$\,keV, and $R_{\rm frac}=-1$
(i.e., only the reflection without the continuum spectrum is shown).
}
\label{fig:xillNS_logne}
\end{figure*}

\begin{figure*}[ht]
\centering
\includegraphics[width=\linewidth,trim={3cm 2cm 3cm 0.5cm}]{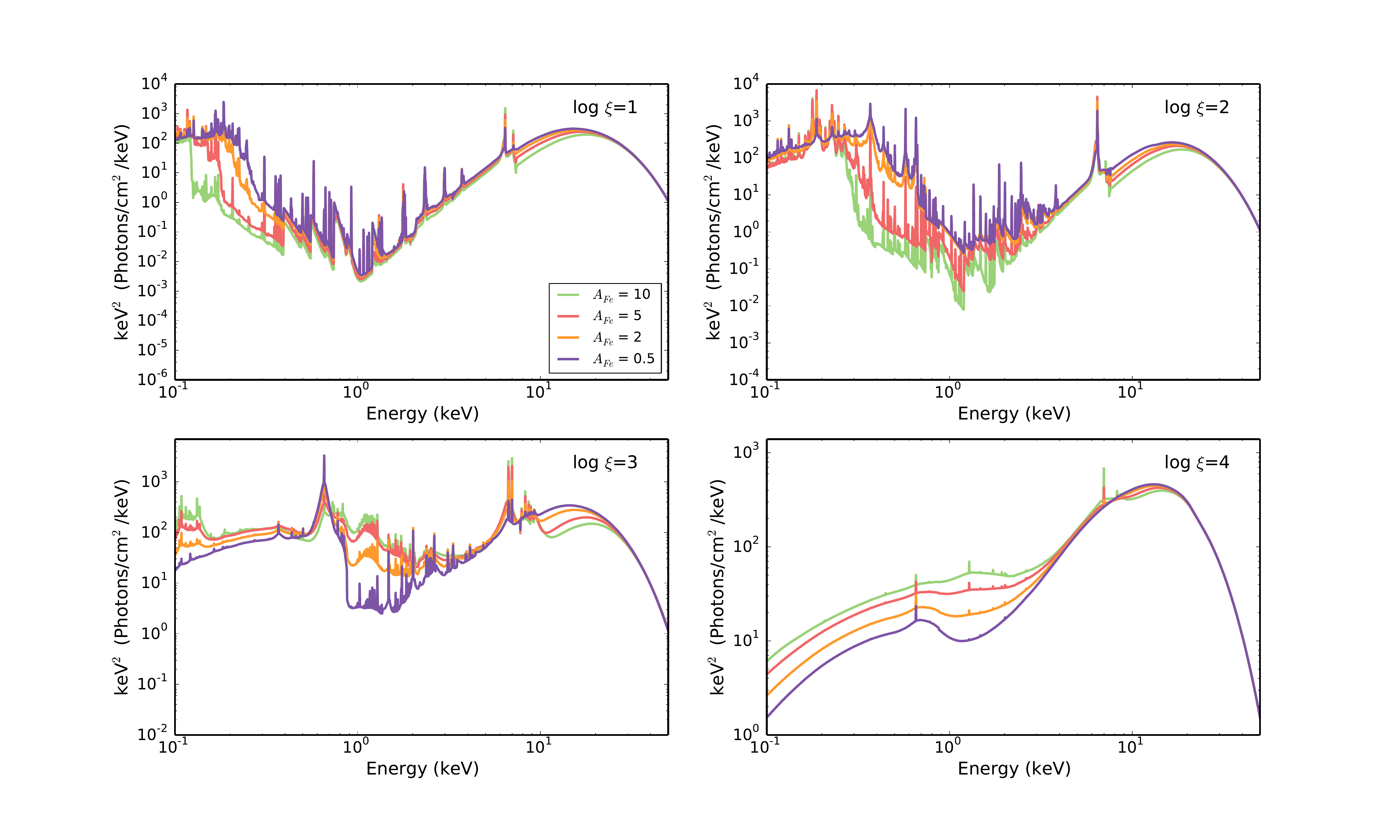}
\caption{
Effects of changing the content of iron for different illuminating fluxes in
the reflection spectra computed with the new \xillverNS\ code.  Each panel
shows the reflected spectra for iron abundance \Afe\ (in Solar units), at a
given ionization parameter \logXi, as indicated. Other parameters common to all
the models shown are: \logne$=19$, $i=30$\deg, \kTbb$=4$\,keV, and $R_{\rm
frac}=-1$ (i.e., only the reflection without the continuum spectrum is shown).
}
\label{fig:xillNS_Afe}
\end{figure*}

Figures~\ref{fig:xillNS_kTbb}, \ref{fig:xillNS_logne}, and \ref{fig:xillNS_Afe}
show examples of the resulting calculations of the reflected spectra with
\xillverNS\, for several combinations of model parameters. The overall behavior
of these models is similar to any of the previous incarnations of the \xillver\
calculations: the continuum of the reflected spectrum follows the general shape
of the incident radiation (a blackbody in this case), with strong departures
caused by a combination of photoelectric absorption, fluorescent emission, and
absorption edges from the different ions in material; as well as the
redistribution of photons due to Compton scattering. 

Specifically, Figure~\ref{fig:xillNS_kTbb} shows the effects of varying the
illuminating radiation field by either changing the net incident flux
(parameterized via $\xi$), or the temperature of the blackbody emission
(\kTbb). As expected, increasing the net flux results in a more ionized slab,
which reduces the amount of spectral features. However, for a given value of
$\xi$, increasing the temperature of the blackbody changes the bulk energy of
the incident photons, producing strong changes in the spectrum. Models with
higher \kTbb\ have more photons in the Fe K band, making the line emission more
efficient, and producing a more noticeable Fe K-edge. Interestingly, for
\logXi$\gtrsim2$, the reflected continuum at low energies ($\lesssim 1$\,keV)
becomes much more prominent (and flatter) than the incident blackbody, due to
the bremsstrahlung (free-free) emissivity.
%

%......................................................................................
%......................................................................................
%

The effect of the increased bremsstrahlung emissivity due to an increment in
the gas density is clearly depicted in Figure~\ref{fig:xillNS_logne}. Each
panel show the comparison of the spectra reflected out slabs with different
densities, but produced with the same illumination spectra. The higher the
density, the more the flux in the reflected continuum at soft energies (see in
particular the case for \logXi$=2$). This behavior resembles closely the
reflection produced with a power-law illumination, as described in
\cite{gar16b}.  At higher densities the free-free heating is enhanced, raising
the temperature of the atmosphere. This increases the amount of broadening in
the spectral features through Compton scattering. The increased density and
temperature may also increase the ionization of some species, which is likely
the reason why the Fe K-edge appears less pronounced (e.g., see \logXi$=1$). At
high ionization (e.g., \logXi$=3$) the changes in the soft flux are less
severe, and in fact, at even higher ionization the effect is inverted. In the
case of \logXi$=4$, the higher density models have the lower flux at $\sim
0.1$\,keV.  This is because the peak of the bremsstrahlung emissivity keeps
shifting to higher energies, likely blending with the incident radiation field. 

Finally, the effects of varying the iron abundance are exemplified in
Figure~\ref{fig:xillNS_Afe}.  As in any of our previous \xillver\ calculations,
the iron abundance has a very predictable effect in the spectrum: any of the
spectral profiles associated with iron, both in emission and absorption, are
enhanced linearly with the increase of \Afe. In particular, the Fe K emission
complex at $\sim 6.4-6.9$\,keV becomes more intense with increased abundance,
as well as the Fe K-edge ($\sim7-9$\,keV) becomes more prominent. This is also
observed at lower energies, where other Fe transitions occur. A good example is
the Fe L-shell emission complex near 1\,keV. In the case of \logXi$=3$, the
reflected flux at those energies increases by almost 2 orders of magnitude,
when comparing the models with \Afe$=0.5$ and \Afe$=10$.

A closer inspection of Figure~\ref{fig:xillNS_Afe} shows an interesting aspect
of these models. At soft energies (below $\sim 1$\,keV), the effects of
increasing the iron abundance seem to invert as the ionization parameter
increases.  At low $\xi$, larger iron abundance increases the photoelectric
opacity causing a drop in the continuum. But at these energies the opacity is
mostly dominated by low-$Z$ ions. However, more iron means that hard X-ray
photons are absorbed more efficiently and at a much lower depth, which makes
the overall ionization of the slab lower. Thus, at low ionization, the larger
the abundance the more neutral the gas appears for the same incident radiation
field. At high values of $\xi$ the situation changes, because most of the ions
are either highly ionized or completely stripped. Thus, increasing the Fe
abundance mostly contributes to increasing the gas temperature through
photoionization heating, which increases the overall ionization state of the
gas. In this regime, models with high iron abundance have the strongest flux at
soft energies.
%

%......................................................................................
%......................................................................................
%

\subsection{Relativistic Reflection}\label{sec:relxillNS}

Similarly to the case of accretion disks around black holes, when the
reflection of X-rays occurs in the regions of the accretion disk closer to the
neutron star, reprocessed photons are affected by light-bending and energy
shifts on their way to the observer. Doppler boosting and gravitational shifts
take place causing a distortion of the spectrum, mainly affecting sharp atomic
features. The magnitude of the spectral blurring increases close to the compact
object. Thus, modeling in detail the spectral shapes provides estimates on the
properties of the accretion disk, including its inner boundary. Under certain
assumptions the inner radius of the disk can then be associated with the radius
of the neutron star, or at the very least it provides a reasonable upper limit
\citep{cac08}.
%

%......................................................................................
\begin{figure*}[ht]
\centering
\includegraphics[width=\linewidth]{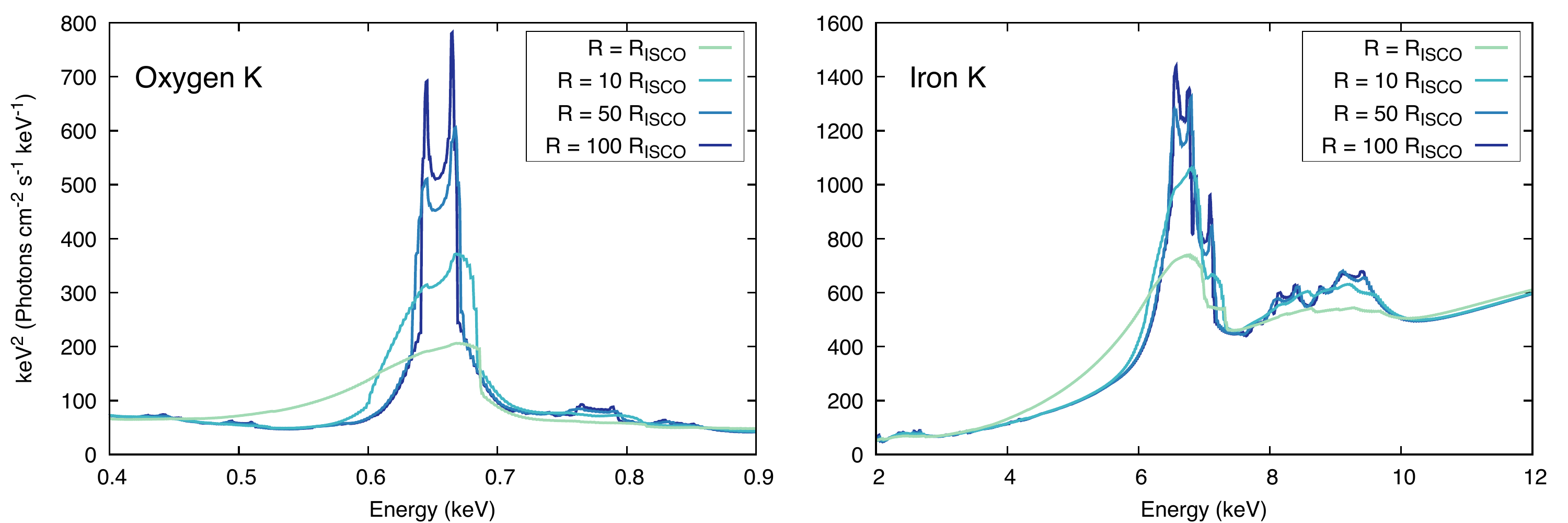}
\caption{Reflection spectra produced with \relxillNS\ for different values of
disk truncation, as indicated. Left and right panels show the emission due to
oxygen and iron K fluorescence, respectively. Other model parameters are:
$q_1=q_2=3$, $a=0$, $i=30$\deg, $R_\mathrm{out}=10^3 R_g$, \kTbb$=4$\,keV,
\logXi$=3.1$, \Afe$=5$, \logne$=15$, and $R_\mathrm{frac}=1$ (i.e., including
both reflection and continuum components). 
}
\label{fig:lines}
\end{figure*}
%......................................................................................
%

Given the importance of properly modeling the relativistic effects in the
reflected component, the reflection models discussed in the previous section
(\xillverNS) have been included in our suite of relativistic reflection models
\relxill\ \citep{dau14,gar14a}. This new flavor of our models, called
\relxillNS, calculates the total spectrum reprocessed in an accretion disk
illuminated by a blackbody radiation field, by integrating the individual
angle-dependent \xillverNS\ reflection spectra emitted by different annuli and
self-consistently including the relativistic effects. We note that all
relativistic effects are computed using the Kerr metric, which correctly
describes the space-time near for black holes and non rotating neutron stars.
However, it is important to point out that deviations can occur from an induced
quadrupole moment as the neutron star becomes oblate in structure as rotation
rate increases. The Kerr metric is a good approximation for the space-time near
a neutron star at low spin values, with a quadrupole-induced deviation of at
most $10$\% from the Kerr metric at $a=0.3$ \citep{sib98}. The discrepancy
becomes larger as break-up of the neutron star is approached
\citep[$\sim25\%$,][]{sib98}, but most neutron stars in LMXBs are expected to
have a spin of $a_*\leq0.3$ \citep{gal08, mil11}.

As in previous models, the emissivity profile of the disk is assumed to follow
a broken power-law profile with inner and outer emissivity indices, and
breaking radius, taken as fit parameters. This is a parameterization of the
illumination pattern rather than a physical model, given the uncertainty on the
exact geometry of the primary source of photons. Meanwhile, \cite{wil18}
studied the illumination of disks around neutron stars using a fully
relativistic ray tracing approach, producing theoretical emissivity profiles
for illumination due to hotspots, bands of emission, and emission by the
entirety of the spherical star surface. In all these cases, the emissivity is
well described by a single power-law with the canonical index slightly steeper
than the canonical value of $-3$.
%

%..................................................................................
%
\begin{deluxetable}{ccc}[ht!]
\tabletypesize{\scriptsize}
\tablecaption{List of Parameters for the \relxillNS\ Model \label{tab:relxillNS}}
\tablecolumns{3}
\tablewidth{0pt}
\tablehead{
\colhead{Parameter} & \colhead{Symbol (Units)} & \colhead{Range}
}
\startdata
Inner Emissivity Index  & $q_1$                   & $[-10,10]$ \\
Outer Emissivity Index  & $q_2$                   & $[-10,10]$ \\
Break Radius            & $R_\mathrm{Br}$ ($R_g$) & $[1-1000]$ \\
Spin Parameter          & $a_*$ ($cJ/GM^2$)       & $[-0.998,0.998]$ \\     
Inclination             & $i$ (degrees)           & $[3,87]$ \\
Inner Disk Radius       & $R_\mathrm{in}$ ($R_\mathrm{ISCO}$) & $[1,1000]$ \\
Outer Disk Radius       & $R_\mathrm{out}$ ($R_g$) & $[1,1000]$ \\
Blackbody Temperature   & \kTbb\ (keV)            & $[0.5,10]$ \\
Ionization Parameter    & \logXi                  & $[1,4]$    \\
Electron Number Density & \logne                  & $[15,19]$  \\
Iron Abundance          & \Afe\ (Solar)           & $[0.5,10]$ \\
Reflection Fraction$^a$ & $R_\mathrm{frac}$       & $[0,10]$ \\
\hline
\enddata
\tablenotetext{a}{If this parameter is set to negative values, the model only 
outputs the reflection component, without the continuum.}
\end{deluxetable}
%
%..................................................................................

In addition to the parameters describing the reflection spectra (see
Table~\ref{tab:xillverNS}), other model parameters include the dimensionless
spin parameter, inclination, inner and outer radius of the disk, and the
reflection fraction. The latter controls the proportion of the blackbody
continuum to the reflection component. Given that the exact origin of the
blackbody emission is unknown, and its geometry is not specified, we
parametrize the emissivity profile as a power-law. Thus, a self-consistent
calculation of the reflection fraction is not possible with this model. For the
same reasons, it is not possible to derive a physical interpretation of the
fitted values. However, the reflection fraction does provides some clues on the
possible geometry of the region responsible for the primary emission, as
relativistic effects strongly affect its value. For example, reflection
dominated spectra are only likely for compact emitting regions, for which
relativistic beaming reduces the flux of the primary component and enhances the
reflection fraction.

The full list of parameters with their allowed ranges is summarized in
Table~\ref{tab:relxillNS}.  Figure~\ref{fig:lines} shows a series of reflected
spectra produced with \relxillNS\ for different values of the inner radius
(i.e., for different degrees of disk truncation), looking in particular at the
emission due to oxygen and iron K transitions. Traditionally, reflection
spectroscopy studies had focused on the Fe K emission given its prominence in
the X-ray spectra.  However, in cases where the continuum is much softer than a
power-law, such as is the case of the blackbody illumination in \relxillNS,
lines from lower-$Z$ elements can also become important probes of relativistic
effects. The oxygen K emission lines are a good example of an alternative
atomic feature that can be used for the estimation of the disk radius and other
model parameters.
%

%----------------------------------------------------------------------------------
\subsection{Comparisons with other reflection models}

In this Section we present a detailed discussion of the performance of the
\xillverNS\ and \relxillNS\ models, and compare them with other reflection
models previously published.

\subsubsection{Fits to the neutron star 4U~1705$-$44}

As a reference source we have chosen the low mass X-ray binary neutron star
system 4U~1705$-$44, which is a well studied bursting source that displays
strong signatures of disk reflection in the X-ray spectrum.  Previous \xmm\
observations taken in the hard and soft states have been analyzed by
\cite{dis09, dai10}, and \cite{egr13}. All these works reported the presence
of iron K-shell emission lines, as well as several other emission features at
softer energies associated with lower-$Z$ elements like argon and calcium. As
discussed in \cite{egr13}, Ar and Ca lines were not included in earlier
reflection codes, until the production of the \xillver\ models.

Although it is
clear that the largest discrepancies between different generations of
reflection models are most significant at softer energies, where the largest
amount of spectral lines are observed, the reflection models featured in this
paper (\xillverNS, \relxillNS) are generally intended for observations from
sources in soft states, where the continuum is dominated by thermal emission
(both from the disk and from the neutron star), which it is expected to
overcome the reflection component at soft energies.
In some systems, however, the accretion disk does not extend down to the
surface of the neutron star, but it is rather truncated at a few gravitational radii,
likely due to the presence of a boundary layer. This causes a fainter disk
emission, allowing for the soft energy features of the reflection spectrum to
be observed. Thus, accurate reflection models with updated atomic data for all
relevant species are necessary to constraint the physical parameters that
describe the accretion flow around neutron stars.

In any case, detailed
comparisons of spectral features at energies below the \nustar\ bandpass
require analysis of data from other observatories, which are prone to vetting
instrumental effects such as photon pile-up \citep[in the case of \xmm;
however, see][]{egr13}, or uncertain calibration (in the case of \nicer). This
would require a much more careful and complicated analysis, far beyond the
scope of the present paper.

We thus restrict our analysis and comparison of all the models to the same
observational dataset: a 29\,ks spectrum in the $3-30$\,keV X-ray band observed
with both Focal Plane Modules onboard of \nustar.  This observation was
previously analyzed by \cite{lud17b} with the \BBrefl\ and \reflionxBB\ models.
The data was reduced using the standard mission procedures as described in
\cite{lud17b}.  In general, the models applied to the \nustar\ observations of
4U~1705$-$44 have the same structure: a multi-temperature disk emission
spectrum modeled with {\tt diskbb} \citep{mit84,mak86}, plus a single
temperature blackbody component (likely originating from the surface or
boundary region of the neutron star), and its corresponding reflection
spectrum. Galactic absorption is modeled with the {\tt TBabs} model assuming
\cite{wil00} abundances and \cite{ver96} cross sections.  Relativistic effects
that distort the reflected radiation produced close to the neutron star are
included via the convolution model \relconv\ \citep{dau10,dau13}.  Thus, in
{\sc xspec} notation the models are written as: %

\begin{itemize}
\item[Model 1:] {\tt TBabs*(diskBB+relconv$\otimes$BBrefl)}
\item[Model 2:] {\tt TBabs*(diskBB+relxillNS)}
\item[Model 3:] {\tt TBabs*(diskBB+BBody+relconv$\otimes$reflionx\_BB)}
\end{itemize}
%

%......................................................................................
\begin{figure*}[ht]
\centering
\includegraphics[width=\linewidth,trim={0 0 0 0}]{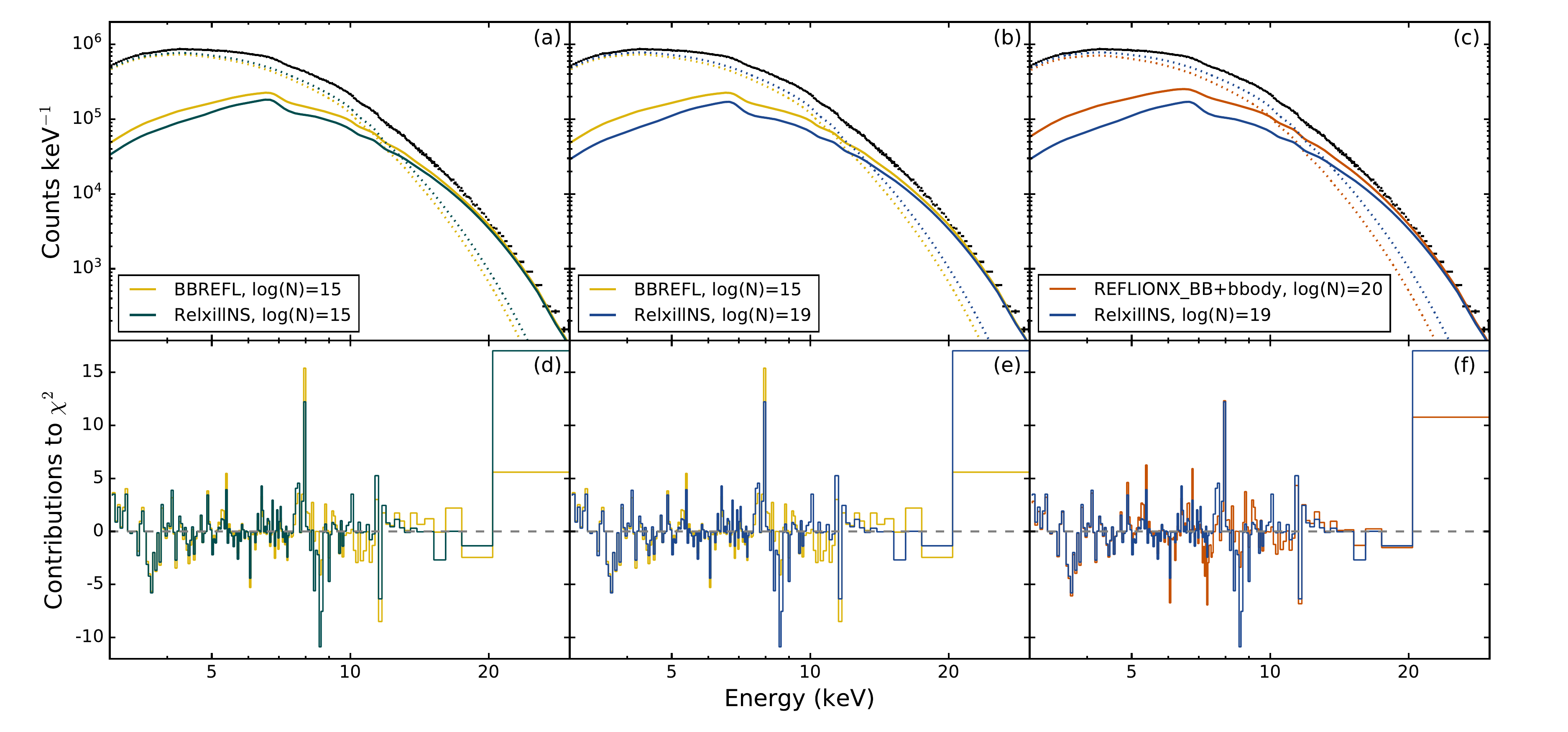}
\caption{
Spectral fits to the \nustar\ data of the neutron star 4U~1705$-$44, using
three different reflection models: \BBrefl, \relxillNS\ and \reflionxBB. Each
column shows a comparison of the fits using \relxillNS\ and \BBrefl\ (left and
middle columns), and \reflionxBB\ (right column). The top row shows the model
components (blackbody continuum, dotted) and reflection spectrum (solid color
lines), and the the total \relxillNS\ model applied to the observational data
(black solid line and data points, respectively). The panels in the bottom row
show the contributions to $\chi^2$ from each fit.
}
\label{fig:comparison}
\end{figure*}
%......................................................................................
%

Here, both \BBrefl\ \citep{bal04} and \reflionxBB\ are based on the same parent
code {\tt reflionx} \citep{ros05}, with the difference they both assume a
blackbody illumination spectrum, rather than the standard power-law. Moreover,
\BBrefl\ assumes local thermo-dynamical equilibrium (LTE) throughout the slab,
while \reflionxBB\ produces a non-LTE calculation that predicts the local flux
at each depth, which accounts for differences in the line profiles.  {\tt
Reflionx\_BB} was constructed as a simple, generally applicable model for
reflection in neutron stars. It uses the standard {\tt reflionx} code, without
additional physics, but replacing the power-law illumination with a black body
spectrum and increasing the density from $10^{15}$\,cm$^{-3}$ to
$10^{20}$\,cm$^{-3}$. It covers a wide range in ionization
($10-10^4$\,erg\,cm\,s$^{-1}$), iron abundance ($0.1-10$ Solar) and black body
temperature ($0.1-10$\,keV).  {\tt Reflionx\_BB} also has a broader range of
elements, charge states, and ionization than {\tt BBRefl}. We have chosen a
version of \BBrefl\ with iron abundance fixed to twice the Solar value, and gas
density of \logne$=15$. Note that \BBrefl\ provides the blackbody continuum,
and thus this component does not need to be explicitly included. The same
applies to \relxillNS, which also self-consistently includes all relativistic
blurring effects, and thus the \relconv\ convolution model is also not
required.
%

%..................................................................................
%
\begin{deluxetable*}{cccccc}[ht!]
\tabletypesize{\scriptsize}
\tablecaption{Comparison of the best-fit parameters to \nustar\ data of 4U~1705$-$44 using
different reflection models \label{tab:comparison}}
\tablecolumns{6}
\tablewidth{0pt}
\tablehead{
\colhead{Component} & \colhead{Parameter} & \colhead{Model~1} & \multicolumn{2}{c}{Model~2}
& \colhead{Model~3}
}
\startdata
{\tt TBabs}     & $N_\mathrm{H} (10^{22}$\,cm$^{-2})$ & $0.7$                  & $0.7$                  & $0.7$                  & $0.7$                  \\
{\tt diskbb}    & $T_\mathrm{in}$\,(keV)              & $2.01^{+0.05}_{-0.04}$ & $2.14^{+0.06}_{-0.06}$ & $2.14^{+0.03}_{-0.04}$ & $1.95^{+0.02}_{-0.04}$ \\
{\tt diskbb}    & $N_\mathrm{d}$                      & $11.4^{+0.6}_{-0.8}$   & $9.5^{+0.8}_{-0.7}$    & $9.7^{+0.5}_{-0.4}$    & $12.5^{+0.9}_{-0.3}$   \\
{\tt BBody}     & $kT$\,(keV)                         & \nodata                & \nodata                & \nodata                & $2.53^{+0.02}_{-0.03}$ \\
{\tt BBody}     & $N_\mathrm{BB}$\,$(10^{-2})$        & \nodata                & \nodata                & \nodata                & $1.21^{+0.06}_{-0.10}$ \\
{\tt relconv}   & $q$                                 & $3.45^{+0.39}_{-0.40}$ & \nodata                & \nodata                & $3.29^{+0.46}_{-0.84}$ \\
{\tt relconv}   & $a_*$\,($cJ/GM^2$)                  & $0.0$                  & \nodata                & \nodata                & $0.0$                  \\
{\tt relconv}   & $i$\,(\deg)                         & $25.5^{+0.9}_{-2.6}$   & \nodata                & \nodata                & $24.7^{+1.2}_{-11.5}$  \\
{\tt relconv}   & $R_\mathrm{in}$\,(ISCO)             & $1.53^{+0.15}_{-0.19}$ & \nodata                & \nodata                & $1.49^{+0.49}_{-0.16}$ \\
{\tt relconv}   & $R_\mathrm{out}$\,($R_g=GM/c^2$)    & $990$                  & \nodata                & \nodata                & $990$                  \\
\BBrefl         & $\log(n_e/$cm$^{-3})$               & $15$                   & \nodata                & \nodata                & \nodata                \\
\BBrefl         & $\log(\xi/$erg\,cm\,s$^{-1})$       & $2.71^{+0.16}_{-0.09}$ & \nodata                & \nodata                & \nodata                \\
\BBrefl         & $kT$\,(keV)                         & $2.67^{+0.05}_{-0.03}$ & \nodata                & \nodata                & \nodata                \\
\BBrefl         & $A_\mathrm{Fe}$\,(Solar$^{\dagger}$)            & $2$                    & \nodata                & \nodata                & \nodata                \\
\BBrefl         & $f_\mathrm{refl}$                   & $0.76^{+0.16}_{-0.07}$ & \nodata                & \nodata                & \nodata                \\
\BBrefl         & $N_\mathrm{BBR}$\,$(10^{-26})$      & $2.2^{+0.4}_{-0.8}$    & \nodata                & \nodata                & \nodata                \\
\reflionxBB     & $\log(n_e/$cm$^{-3})$               & \nodata                & \nodata                & \nodata                & $20$                   \\
\reflionxBB     & $\xi$\,(erg\,cm\,s$^{-1}$)          & \nodata                & \nodata                & \nodata                & $407\pm218$            \\
\reflionxBB     & $A_\mathrm{Fe}$\,(Solar$^{\dagger}$)            & \nodata                & \nodata                & \nodata                & $2.7^{+1.4}_{-1.7}$    \\
\reflionxBB     & $N_\mathrm{RX}$                     & \nodata                & \nodata                & \nodata                & $0.57\pm0.14$          \\
\relxillNS      & $q$                                 & \nodata                & $3.7^{+0.7}_{-0.3}$    & $3.9^{+0.8}_{-0.4}$    & \nodata                \\
\relxillNS      & $a_*$\,($cJ/GM^2$)                  & \nodata                & $0$                    & $0$                    & \nodata                \\
\relxillNS      & $i$\,(\deg)                         & \nodata                & $30.5^{+1.0}_{-1.1}$   & $30.9^{+1.0}_{-1.0}$   & \nodata                \\
\relxillNS      & $R_\mathrm{in}$\,(ISCO)             & \nodata                & $1.74^{+0.18}_{-0.22}$ & $1.64^{+0.18}_{-0.22}$ & \nodata                \\
\relxillNS      & $R_\mathrm{out}$\,($R_g=GM/c^2$)    & \nodata                & $990$                  & $990$                  & \nodata                \\
\relxillNS      & $kT_\mathrm{BB}$\,(keV)             & \nodata                & $2.80^{+0.13}_{-0.08}$ & $2.90^{+0.09}_{-0.07}$ & \nodata                \\
\relxillNS      & $\log(n_e/$cm$^{-3})$               & \nodata                & $15$                   & $19$                   & \nodata                \\
\relxillNS      & $\log(\xi/$erg\,cm\,s$^{-1})$       & \nodata                & $2.64^{+0.07}_{-0.04}$ & $2.56^{+0.02}_{-0.01}$ & \nodata                \\
\relxillNS      & $A_\mathrm{Fe}$\,(Solar$^{\dagger}$)            & \nodata                & $2.63$                 & $2.63$                 & \nodata                \\
\relxillNS      & $R_f$                               & \nodata                & $0.9^{+0.7}_{-0.3}$    & $1.3^{+0.6}_{-0.4}$    & \nodata                \\
\relxillNS      & $N_\mathrm{XNS}$\,($10^{-4}$)       & \nodata                & $4.6^{+2.0}_{-1.9}$    & $3.3^{+1.1}_{-1.0}$    & \nodata                \\
\hline
                & $\chi^2$                            & $797.59$               & $801.50$               & $789.91$               & $809.16$               \\
                & $\nu$                               & $665$                  & $665$                  & $665$                  & $664$                  \\
                & $\chi_{\nu}^2$                      & $1.20$                 & $1.21$                 & $1.19$                 & $1.22$                 \\
\enddata
\tablenotetext{}{Parameters with no uncertainties were fixed at the quoted value.}
\tablenotetext{\dagger}{The solar abundances in the \BBrefl\ and \reflionxBB\
models are from \cite{mor83}, while \xillverNS\ and \relxillNS\ use the
standard from \cite{gre98}, which quotes an abundance for iron $\sim 30$\% lower
\citep[see Table~1 in][]{gar13a}. Thus, $A_\mathrm{Fe}=2$ in \BBrefl\ its
equivalent to $A_\mathrm{Fe}=2.63$ in \relxillNS.
}
\end{deluxetable*}
%
%..................................................................................

The best-fit parameters and statistics for the goodness of the fits are
summarized in Table~\ref{tab:comparison}.  In all the cases, we fixed the
hydrogen column density to the \cite{kal05} value, and the spin parameter to
zero.  The disk density in \relxillNS\ (Model~2) was also fixed to match as
close as possible the values assumed in \BBrefl\ ($10^{15}$\,cm$^{-3}$) and
\reflionxBB\ ($10^{20}$\,cm$^{-3}$).  Thus, for Model~2 we present results
fixing the density to $10^{15}$ and $10^{19}$\,cm$^{-3}$ (the latter being the
maximum value possible in the current version of our model).  The results show
that the density has a minor effect on the quality of the fits. The majority of
the model parameters are consistent among these fits: importantly, the
blackbody temperature, inner disk radius, and ionization parameter are all in
agreement within their uncertainty levels. Iron abundance is in good agreement
between Models 1 and 2, when the differences in the Solar value are considered
(see caption in Table~\ref{tab:comparison}).  The fits with \relxillNS\
(Model~2) yield disk inclinations a few degrees higher than the other two
models, likely due to differences in the atomic data implemented by each code,
which can affect the details of the Fe K emission profile.  In general, the
three models provide fits of comparable quality, with only a marginal
preference of \relxillNS\ (Model 2) using the highest density value
(\logne$=19$), with a $\chi^2$ of 790 compared to 798 of \BBrefl\ (both for 665
degrees of freedom, d.o.f.), and 809 for \reflionxBB\ for 664 d.o.f. 

A comparison of the performance of the different fits in reproducing the
\nustar\ data in the $3-30$\,keV region is shown in
Figure~\ref{fig:comparison}, where we include the model components (incident
blackbody and reflection spectra), the total model with the data, and the
residuals of each fit. We compare the \relxillNS\ fit (Model~2) with low and
high density against the \BBrefl\ fit (Model~1), and against the \reflionxBB\
fit (Model~3) for the high density case.  In all cases the strongest residuals
are seen at high energies, above $\sim20$\,keV, where the source counts start
to get dominated by the background. An additional narrow emission feature is
observed at $\sim8$\,keV, as well as some absorption at $\sim8.5$\,keV, whose
origin is unclear. However, we emphasize that the analysis presented here is
intended for comparative purposes only, and thus a more detailed examination of
the physics of this system is left for future publications. The residuals show
otherwise a relatively similar fit by all the models, as also demonstrated by
the fit statistics.
%

%......................................................................................
\begin{figure}
\centering
\includegraphics[width=\linewidth]{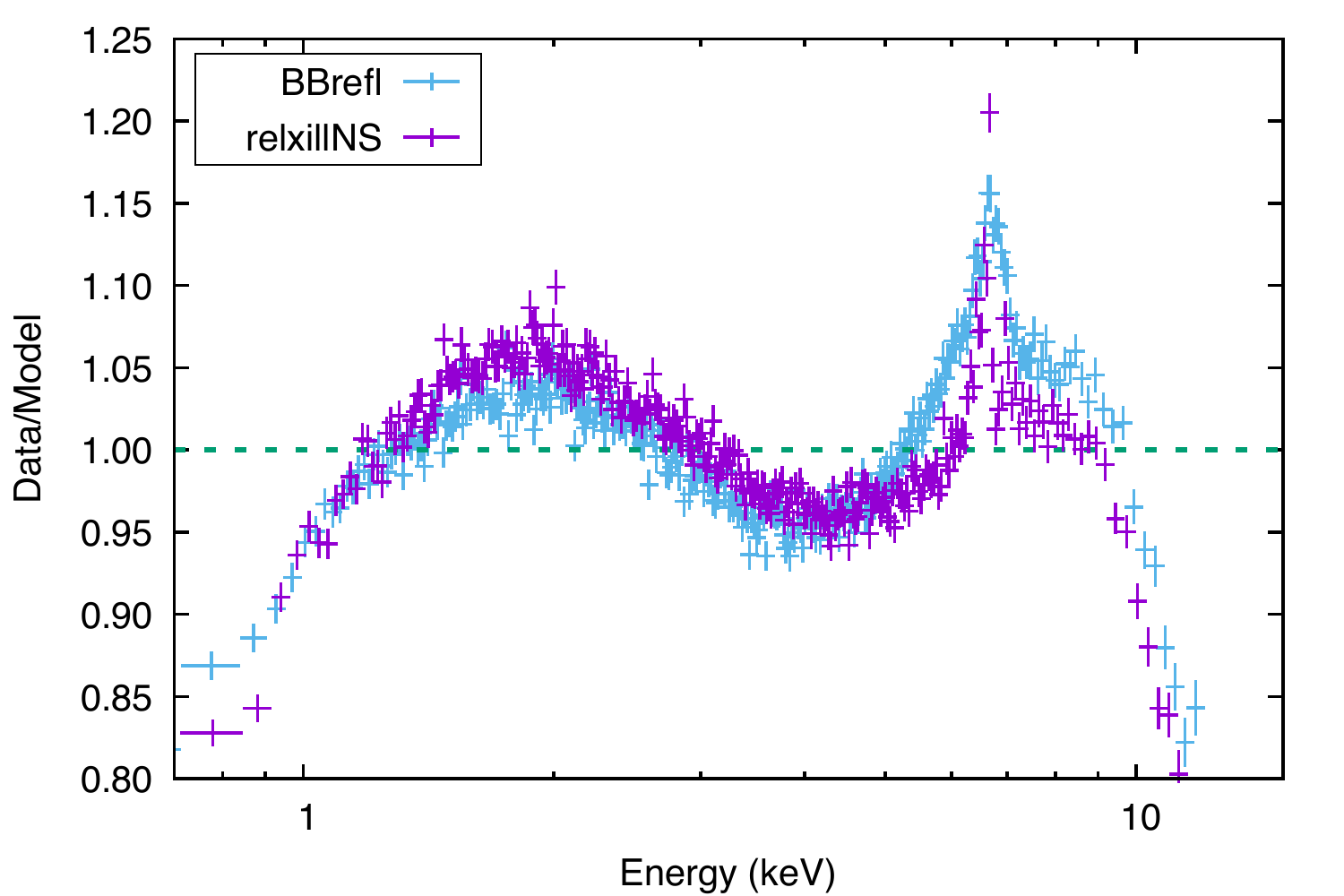}
\caption{
Ratio plot of simulated {\it XRISM-Resolve} observations to an absorbed
blackbody plus power-law model. The simulations were produced using the
best-fit parameters for Model~1 (based on \BBrefl) and Model~2 (based on
\relxillNS). In both cases we use the {\it Resolve} response matrix for a 5\,eV
spectral resolution at 6\,keV, and a exposure time of 20\,ks.
}
\label{fig:xrism}
\end{figure}
%......................................................................................
%

\subsubsection{XRISM Simulations}

Although the fits discussed in the previous Section yielded results with very
similar statistical quality, the different models implemented are not
identical. It is thus expected that the new generation of X-ray observatories
with improved effective area and spectral resolution will provide data with
sufficient signal to distinguish between these models.  To demonstrate this, we
have carried out simulations of observational data using the instrumental
response for the {\it Resolve} instrument onboard of the {\it X-Ray Imaging and
Spectroscopy} mission ({\it XRISM}) \citep{tas18}. {\it Resolve} is a soft
X-ray micro-calorimeter spectrometer, which provides non-dispersive $5-8$\,eV
energy resolution in the $0.3-12$\,keV bandpass.

Specifically for these simulations, we have used the most optimistic RMFs for
5\,eV resolution together with the ARF files for the case of gate-valve open
(i.e., no filter). The simulations were produced with the {\tt fakeit} routine
in {\sc xspec}, based on the best-fits produced with Model~1 (which is based on
\BBrefl) and Model~2 (based on \relxillNS), for the \nustar\ data of
4U~1705$-$44 shown in the previous Section. The flux predicted by this fit is
close to 100\,mCrab in the $2-10$\,keV band, and we assumed a 20\,ks exposure
time for each simulation.

Figure~\ref{fig:xrism} shows the two {\it XRISM-Resolve} simulations as a ratio plot
of the data to a model for the continuum, based on an absorbed blackbody plus power-law
model. We note that the power-law is included to provide a continuum that better resembles
the observations and improve clarity in the ratio plot. The superior energy resolution
of the micro-calorimeter makes evident the differences between these two models, particularly
in the shape of Fe K emission, as well as the overall shape of the continuum at softer 
energies. Interestingly, \relxillNS\ predicts a narrower Fe K profile than \BBrefl.
Given that the two models were evaluated for similar parameters (e.g., ionization, spin,
inner radius, inclination), the difference in the width of the emission
cannot be due to a different relativistic smearing, but rather to the intrinsic
shape of the reflection spectrum in the local frame. When relativistic effects
are excluded, the most significant source of line broadening is due to
Comptonization of reflected photons in the hot layers of the disk's atmosphere.
Photons are shifted to higher or lower energies due to electron scattering,
depending on the optical depth and temperature of the material. In fact, for
high enough temperatures ($T\sim 10^7-10^8$\,K), the Compton kernel is dominated by
the contribution of the kinetic energy of the electrons, causing a rather
symmetric broadening of spectral features \citep{gar20}. The differences
observed between \relxillNS\ and \BBrefl\ suggests a dissimilar solution
for the state of the gas between the two models. Our simulations show that such
discrepancies can only be distinguished with the superior energy resolution of
future instruments. Conversely,
this simple comparison demonstrates the importance of accurate models for
the upcoming generation of X-ray instruments.
%

%......................................................................................
\begin{figure*}
\centering
\includegraphics[width=\linewidth]{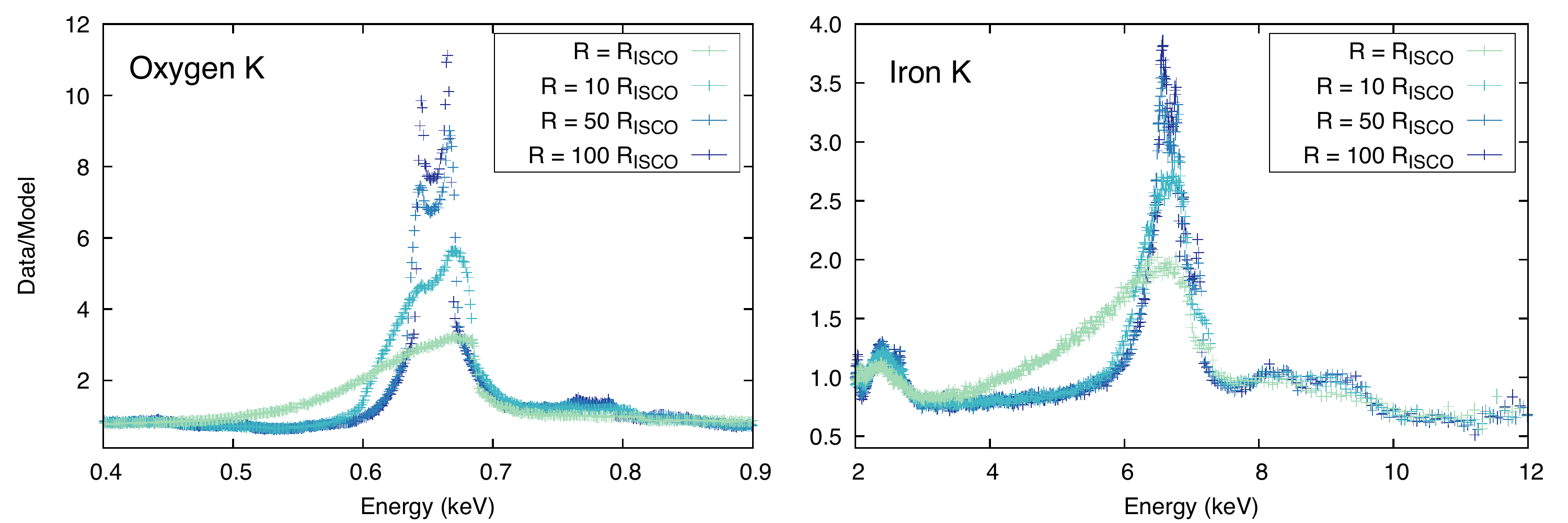}
\caption{
Ratio plot of {\it Athena} X-IFU simulations to an absorbed blackbody plus power-law
model. The simulations were generated with \relxillNS\ for different values of
disk truncation, using the same parameters of the models in Figure~\ref{fig:lines}.
In all the cases, the exposure time is 10\,ks for a flux source of 10\,mCrab.
}
\label{fig:athena}
\end{figure*}
%......................................................................................
%

\subsubsection{Athena Simulations}

In order to further emphasize the diagnostic power of our new reflection models
in combination with the enhanced instrumental capabilities of planned
facilities, we have also carried out simulations for the flagship mission {\it
Athena} \citep[{\it Advanced Telescope for High Energy Astrophysics};][]{nan13},
currently under development and planned to be operational in the mid 2030's.
The {\it X-ray Integral Field Unit} (X-IFU) is {\it Athena}'s micro-calorimeter
spectrometer, which is expected to deliver data with an unprecedented energy
resolution of 2.5\,eV or better up to 7\,keV \citep{bar18}.

In this case we carried out simulations based on the \relxillNS\ models
presented in Section~\ref{sec:relxillNS} (shown in Figure~\ref{fig:lines}),
i.e., $q_1=q_2=3$, $a=0$, $i=30$\deg, $R_\mathrm{out}=10^3 R_g$,
\kTbb$=4$\,keV, \logXi$=3.1$, \Afe$=5$, \logne$=15$, and $R_\mathrm{frac}=1$
(including both reflection and continuum components). The four different
cases shown correspond to models with disk inner radius set at $1, 10, 50$, and
$100 R_\mathrm{ISCO}$, as indicated.  For all these cases Galactic absorption
is included via the {\tt TBabs} model assuming a low H column density of $N_H =
10^{20}$\,cm$^{-2}$, and a source flux of 10\,mCrab. Sources with
fluxes of $\sim 100$\,mCrab or more will require a defocussed mode (which
degrades the energy resolution), in order to prevent photon pile-up. However,
owing to the significantly larger effective area of {\it Athena}, observations
of moderately bright source (10\,mCrab) with a exposure of 10\,ks will provide
sufficient signal-to-noise to resolve most of the structure of relativistically
broadened atomic lines.

The X-IFU simulations are shown in Figure~\ref{fig:athena}, which displays the
same energy ranges depicted in Figure~\ref{fig:lines} for the O and Fe K-shell transitions.
These simulations clearly show that with relatively short observations precise 
constraints on important parameters like the disk inner radius will be easily
achievable. The details of the line emission are almost fully resolved, particularly
in the case of the O K emission, where the double horn of the line profile is
clearly resolved at even large truncation radii. However, we note that the 
conditions for the present simulations are overly optimistic: we have chosen 
to simulate the case of a source with very low Galactic absorption with no 
disk emission (only the blackbody continuum and its corresponding reflection
components are included), in order to emphasize the strength of the O K emission.
In reality, most of the sources for which \relxillNS\ was designed are expected
to show a strong disk component which will likely outshine the reflection emission
at soft energies, making the detection of oxygen lines challenging. Nevertheless,
this example demonstrates the potential capabilities of future instruments in
resolving the detailed structure of the spectral profiles.

\subsubsection{Comparison with convolution models}

\begin{figure*}[ht]
\centering
\includegraphics[width=\linewidth]{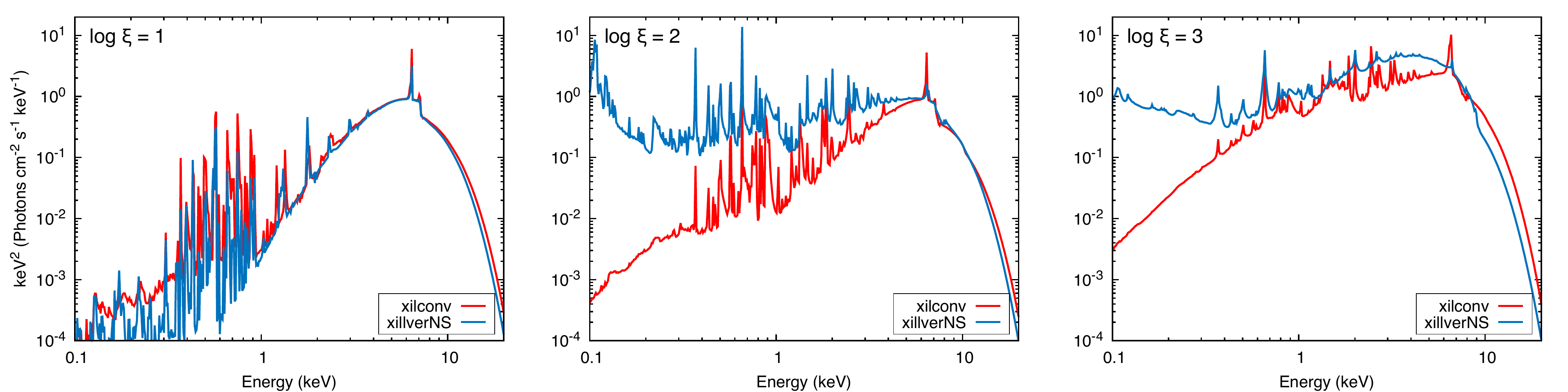}
\caption{
Comparison of \xillverNS\ spectra with those predicted by the {\tt
xilconv$\otimes$bbody} model, for three different values of the ionization
parameter \logXi, as indicated. For all these cases the other model parameters
are: \kTbb$=1$\,keV, \Afe=1, \logne$=15$, $i=60$\,\deg, and $R_{\rm frac}=-1$.
The \xillverNS\ spectra were re-normalized such that the input blackbody
spectrum in the two models are the same when $R_{\rm frac}=0$.
}
\label{fig:xilconv}
\end{figure*}

Another model used for the analysis of neutron star X-ray spectra is \xilconv,
which is an updated version of the \rfxconv\ model \citep{kol11}, as first
described in \cite{don06}. The only difference between these two models is that
\rfxconv\ uses the reflection tables produced with the \reflionx\ code, while
\xilconv\ uses those from \xillver. These models act as a convolution kernel
upon any spectral component, allowing the user to input any desired continuum
spectrum. The model then determines the average power-law index in the
$2-10$\,keV region, and selects the corresponding reflection spectra from
the precomputed table. The main caveat of this procedure is the lack of
self-consistency between the input spectrum and the reflection, as the latter
is always chosen from the standard \xillver\ library computed with the
illumination in form of a power-law, regardless of what input spectrum is
selected. This model is thus prone to spurious results while fitting
observational data. This is demonstrated in Figure~\ref{fig:xilconv}, where we
compare the \xillverNS\ spectra with those predicted by a {\tt
xilconv$\otimes$bbody} model with the same input parameters. We found that for
low ionization parameters the two models compare reasonably well. However, at
higher ionization the discrepancy is striking: the overall shape of the
reflection spectrum is dramatically different at all energies, with \xilconv\
consistently under-predicting the flux at soft energies, and over-predicting
the Fe K emission. We thus advise against the use of these models, particularly
in cases where the input spectrum is different from the standard power-law shape.

%......................................................................................
%......................................................................................
%

%----------------------------------------------------------------------------------
\vspace{1cm}
\section{Discussion \& Conclusions}\label{sec:discon}

We have presented new X-ray reflection models for geometrically-thin,
optically-thick accretion disks around compact objects. The illumination
spectrum is assumed to be a blackbody emission at a given temperature, which
after being reprocessed by the disk, produces a reflection spectrum with
characteristic signatures of great diagnostic potential.  The new reflection
models featured here, \xillverNS, and its relativistic counterpart \relxillNS,
are primarily intended for the interpretation of the X-ray spectrum observed
from accretion disks around neutron stars, which are now commonly observed with
current X-ray satellites. The superior capabilities of the \xmm, \nicer\ and
\nustar\ observatories have opened new venues to study these systems, and thus
accurate X-ray reflection models are crucial for the interpretation of new
observational data.  Similar reflection models have been produced in the past
with outdated codes and atomic databases \citep[e.g.][]{bal04,ros07}. Here, we
provide a major update by implementing angle-resolved radiative transfer
calculations which make use of the most up-to-date collection of atomic
parameters for inner-shell transitions.

A simple test case to the \nustar\ data of 4U~1705$-$44 shows a satisfactory
performance of the new \relxillNS\ in describing the observational data.
Comparisons of fits with earlier models like \BBrefl\ and \reflionxBB\ show
very good consistency between the models.  Although the newer atomic database
included in \relxillNS\ is expected to have a more significant impact at softer
energies, we defer such a detailed analysis for future publications. Meanwhile,
simulations of observations with new generation of X-ray observatories such as
{\it XRISM-Resolve} and {\it Athena} X-IFU revealed the large diagnostic
potential of these new models in describing the detailed structure of the
reprocessed emission, which will be required to describe the high-resolution
spectrum delivered by upcoming micro-calorimeter instruments.

Preliminary versions of the \xillverNS\ and \relxillNS\ models shown in this
paper have already been implemented in several previous works. The model has
been tested and used on a \nicer\ observation of Serpens~X-1 \citep{lud18}, a
\nustar\ observation of GX~3+1 \citep{lud19}, and a joint \nicer\ and \nustar\
observation of 4U~1735$-$44 \citep{lud20}. These are all persistently accreting
``atoll" sources \citep[named after characteristic island-like patterns that
are traced out in hardness-intensity and color-color diagrams;][]{has89}.

In all spectral fits, the models have been able to successfully describe the
data, providing evidence for relativistically smeared atomic lines.
The density of the inner disk was inferred to be higher than the
$10^{15}$~cm$^{-3}$ that used to be the standard value assumed in reflection
models. The emissivity index is less than 4, which is consistent with
expectations for a disk illuminated by a neutron star \citep{wil18}. The limits
on inner disk radius constrained via \relxillNS\ allowed for limits to be
placed on the dipolar magnetospheric strengths of the neutron stars and
presence of boundary layer regions in these systems. Serpens~X-1, in
particular, was a useful test since this source exhibited multiple broad
emission features (i.e., Fe~L and Fe~K). Through employing the \relxillNS\
model it became clear that the Fe~L emission profile was more complex than a
single line, but rather a blend of emission from lower-$Z$ elements, such as
Mg~{\sc iii-vii} \citep[see Figure~4 in][]{lud18}.

Unexpectedly, these models have also played a fundamental role in the detection
of returning disk radiation. This is a general relativistic effect
theoretically predicted by \cite{cun76}, in which thermal disk photons are
returned to the other regions of the disk due to the strong gravitational
bending. The first observational evidences for returning radiation was recently
found in soft-state observations of several black hole binaries, such as
\jj1550\ \citep{con20}, 4U~1630$-$47 \citep{con21}, EXO~1846-031 \citep{wan21},
and MAXI~J0637$-$430 \citep{laz21}.  In all these works, the \relxillNS\ model
was implemented in the spectral fits as a proxy for reflection produced by
returning disk radiation.

We note that the present setup of \relxillNS\ is somewhat simplistic, as it
only considers illumination with a single temperature thermal emitter.  Future
versions of these models will likely be expanded to also include non-thermal
illumination (i.e., power-law) in combination with the blackbody emission, in
order to closely resemble the X-ray continuum observed in several neutron star
systems such as Serpens~X-1 \citep{cac08}, 4U~1728$-$34 \citep{sle16},
Aquila~X-1 \citep{lud17}, XTE~J1709$-$267 \citep{lud17c}, and 4U~1543$-$624
\citep{lud19b}, among several others. Modern observations have shown that the
power-law continuum is best modelled with a thermal Comptonization model
\citep{mat17}, just like in the case of black hole binaries. We thus plan to
implement the newest version of the {\tt thComp} mode \citep{zdz20} in upcoming
version of \relxillNS, in close similarity to the \relxillCp\ family of models
(note, however, that in \relxillCp\ there is not thermal component in the
illuminating spectrum).

Implementing a Comptonization continuum to fit neutron stars will also require
the possibility for a variable photon seed temperature, that can be set well
above the fixed value of $0.05$\,keV in the current \relxillCp\ models. This will be
particularly important for sources in the soft state, where the disk
temperature can reach the $0.5-1$\,keV range. Such a model feature will also be
useful to correctly reproduce sources in the intermediate state (especially
atoll sources), which display an X-ray continuum departing from a simple
blackbody spectrum.

A relatively weak and steep power-law in addition to the thermal components can
also be observed during the soft states of Z and atoll sources, with photon
indices larger than 2 and an overall flux of no more than $10\%$ of the total
emission \citep[see,][]{dis00,dam01,pir07,pin16}.
However, in these cases the contribution of the power-law component to the
illuminating spectrum is weak and will no likely affect the reflection spectrum
in a significant manner.

Another limitation of the models presented here is due to the fact that it uses
the Kerr metric to describe the space-time near the compact object.
Technically, this metric is correct for black holes, and sufficiently accurate
for non rotating neutron stars, but it becomes increasingly inaccurate for
larger spins. The exact metric near a neutron star depends on its mass and
equation of state, which are largely unknown. We thus caution users to proceed
with care when allowing non-zero spins while fitting data neutron star systems,
and we generally recommend no to exceed spin values above $a_*=0.3$.  Both the
new reflection model flavor \relxillNS, and its non-relativistic counterpart
\xillverNS, are publicly distributed to the community in our suite of models
{\sc relxill}, {\tt
v-1.5.0}\footnote{\url{https://www.sternwarte.uni-erlangen.de/research/relxill}}.
%

%==================================================================================
%
%
\acknowledgments 
We thank the referee for thoughtful comments that greatly improved this paper.
We also thank E. Kara, D. Barret and E. Kammoun for clarifications regarding
the simulations with {\it XRISM} and {\it Athena} instrumental responses.
J.A.G. acknowledges support from NASA ATP grant 80NSSC20K0540, 
from the Smithsonian Astrophysical Observatory grant AR0-21003X, and from the
Alexander von Humboldt Foundation.  R.M.L. acknowledges the support of NASA
through the Hubble Fellowship Program grant HST-HF2-51440.001.  This work was
partially supported under NASA contract No.  NNG08FD60C and made use of data
from the NuSTAR mission, a project led by the California Institute of
Technology, managed by the Jet Propulsion Laboratory, and funded by the
National Aeronautics and Space Administration. We thank the NuSTAR Operations,
Software, and Calibration teams for support with the execution and analysis of
these observations. This research has made use of the NuSTAR Data Analysis
Software (NuSTARDAS), jointly developed by the ASI Science Data Center (ASDC,
Italy) and the California Institute of Technology (USA).

\vspace{5mm}
\facilities{NuSTAR}

\vspace{5mm}
\software{{\sc xspec} \citep[v12.10.0c;][]{arn96},
{\sc xillver} \citep{gar10,gar13a}, {\sc relxill}
\citep[v1.5.0;][]{gar14a,dau14}.}

%
%
%==============================================================================
%
%
\bibliographystyle{aasjournal}
\bibliography{my-references}
%
%==================================================================================
%
%
%
\end{document}